\documentclass[12pt]{article}
\usepackage{graphicx}
        \usepackage{amssymb}
        \usepackage{amsmath}
\def \d{\partial}
\def \bv{{\bf v}}

\def \bomega{\boldsymbol{\omega}}

\begin{document}

\title{Lagrangian structure functions in fully-developed hydrodynamical
turbulence}

\author{ K.P. Zybin\footnote{E-mail: zybin@lpi.ru}, V.A. Sirota, A.S. Ilyin, A.V. Gurevich
\\
\small P.N.Lebedev Physical Institute, Russian Academy of Sciences,
Moscow, Russia}

\date{}

\maketitle

\begin{abstract}
The Lagrangian velocity structure functions in the inertial range
of fully
developed fluid turbulence are derived 
basing on the Navier-Stokes equations.  For time $\tau$ much
smaller than the correlation time, the structure functions are
shown to obey the scaling relations $K_n(\tau)\propto
\tau^{\zeta_n}$. The scaling exponents $\zeta_n$ are calculated
analytically. The obtained values are in amazing agreement with
the unique  experimental results of the Bodenschatz group
\cite{Bod2}. New notion -- the Lagrangian position structure
functions $R_n(\tau)$ is introduced.
 All the $R_n$ of the order  $n>3$ are shown to have a universal scaling.

\end{abstract}

\section{Introduction}
The spectrum of velocity pulsations  in a turbulent flow can be
naturally divided into three ranges: the large scales,  the
inertial range, and the viscous range. The largest scales $L$ are
comparable with the size of the turbulent volume. These are,
roughly speaking, the scales of the largest vortices generated in
the flow. The large-scale turbulence is a direct result of laminar
flow instability for given boundary conditions. It is always
non-uniform and anisotropic, the amplitudes of velocity pulsations
are the largest at these scales. The most part of turbulent energy
is contained in the large scales. On the other hand, viscous
effects play a dominant role at scales smaller than the Kolmogorov
scale $\eta=L/{R}^{3/4}$, where $R$ is the Reynolds number
\cite{Landau}.  As the viscosity decreases, the dissipative range becomes
narrower.
The intermediate range of scales $\eta \ll \lambda
\ll L$ is called the inertial range. In these scales, there is no
energy dissipation; the energy flows in Fourrier space from large
scales to the viscous range. Since viscosity is negligible in the
inertial range, the Navier-Stokes equation  can be reduced to the
incompressible Euler limit \cite{Frisch}.

Hereafter we discuss the inertial range of locally homogeneous,
locally isotropic, statistically stationary turbulent flow of
incompressible liquid.
 The phenomenological theory of such turbulence constructed by
Kolmogorov (K41 theory) predicts universal scaling laws for all
statistical values. However,  experiments have revealed
departure from K41 caused by intermittency.

In \cite{ZSIG} we proposed a new model of turbulence explaining
the intermittency and built on the Navier-Stokes Equation. Our idea was that the main
role in statistics (in particular, in
structure functions) belongs to 
the regions with very high
vorticity. 
We showed that these regions stretch out taking the form of vortex
filaments. We also showed that the growth of vorticity in the
filaments is caused by large-scale pulsations. 
A considerable simplification of the problem is reached by
treating the fully developed turbulence  from Lagrangian
viewpoint, which is natural from theoretical point of view
\cite{Yeung}. We obtained the equation for the probability
distribution function (PDF) of vorticity, and time dependence of
its statistical moments. Note that recently the model equation
interpreting the experimental data have been proposed \cite{Beck}.
This model is phenomenological and is not based on the
Navier-Stokes Equation.

This paper is a continuation of \cite{ZSIG}. We apply the theory
developed in \cite{ZSIG} to obtain the Lagrangian statistical
moments of velocity differences also called the Lagrangian
structure functions
\begin{equation} \label{Kn}
K_n (\tau) =\langle |{\bf v}(t+\tau) - {\bf v}(t)|^n \rangle \ ,
\end{equation}
where the values of velocity are taken along the trajectory of
a liquid particle, and the angle brackets denote the  average over
the ensemble of trajectories. We find a relation connecting the
Lagrangian structure functions of different orders. It may be
used as an 
analogue to the extended self-similarity ansatz (established for
the Euler case) \cite{Benzi,Bod2} in the Lagrangian case.

Recently there has been a significant experimental progress in
measuring the Lagrangian statistical properties  \cite{Bod2,
Bod1}.  We compare the predictions of our  theory with  recent
experimental results of the Bodenschatz group \cite{Bod2} and find
a wonderful agreement.

The paper is organized as follows. In \S 2, we recall briefly the
method and some of the results obtained in \cite{ZSIG}. In
particular, we derive the time dependence of statistical moments
of vorticity in a vortex filament. In \S 3, basing on these
results, we calculate the scaling exponents of the Lagrangian
velocity structure functions (\ref{Kn}). We also derive a relation
between the structure functions of different orders, which may be
valid even in the cases when there is no scaling (for example,
near to the boundaries of the inertial interval).  
In \S 4, the theoretical results are compared with the
experimental data.  In \S 5, we introduce a new 
notion: Lagrangian structure functions of position. We
calculate their scaling exponents  on the basis of our model.
These  results can also be checked by experiments.

\section{ Properties of turbulent structures in the inertial interval. }

We consider the Navier-Stokes equation for incompressible liquid.
At the scales  
inside the inertial range the equation takes the form of the Euler
equation:
\begin{equation} \label{hd}
\frac{\d \bv }{\d t} + (\bv \cdot \nabla ) {\bv} + \frac{\nabla
p}{\rho} = 0 \,, \qquad \nabla \cdot \bv =0 \ ,
\end{equation}
where  $\bv$ is the velocity of the flow and $p$ is the  pressure.
The density $\rho$ is taken to be unity below. The second equation
expresses the incompressibility of the liquid. The restriction of
scales by the inertial range presumes, in particular, that we
consider only smooth initial conditions. The equation for pressure
follows from (\ref{hd}):
\begin{equation} \label{deltaP}
-\Delta p = \nabla _i v _j \cdot \nabla _j v _i
\end{equation}
Thus, the equations system (\ref{hd}) is complete. We note that
(\ref{hd})can be rewritten as
$$   
\frac{\partial{\bomega}}{\partial t} = \mbox{rot}\,[{\bf
v},{\bomega}] \ ,
$$
where ${\bomega }= \mbox{ rot}\,{\bf v}$ is vorticity.

We now discuss the formation of stretched structures -- vortex
filaments -- in the turbulent flow. 
They appear as a result of flow instability  caused by large-scale
pressure pulsations \cite{ZSIG}. Random large-scale forces stretch
a liquid drop forming a filament. The main role in the process
belongs to the incompressibility of the liquid. First, the volume
conservation results in transversal compression of the filament
during its straining. Conservation of angular momentum then leads
to acceleration of rotating and hence to growth of vorticity in
the filament. Second, the sound speed in an incompressible liquid
is infinite. This means an instant transport of large-scale
pressure pulsations, which cause the stretching of the filaments.

To illustrate these statements, we consider 
an axially symmetric flow. The hydrodynamic equations for the
radial, azimuthal, and axial velocity components $v_r$,
$v_{\phi}$, and $v_z$ are
\begin{eqnarray} \frac{\partial v_r}{\partial t} + v_r
\frac{\partial v_r}{\partial r} + v_z \frac{\partial v_r}{\partial
z} -\frac{v_{\phi}^2}{r} = -\frac{\partial p}{\partial r}
\nonumber
\\
\frac{\partial v_{\phi}}{\partial t} + v_r \frac{\partial
v_{\phi}}{\partial r} + v_z \frac{\partial v_{\phi}}{\partial z}
+\frac{v_{\phi}v_r}{r} = 0 \label{cylindr} \\  \nonumber
 \frac{\partial v_z}{\partial t} + v_r
\frac{\partial v_z}{\partial r} + v_z \frac{\partial v_z}{\partial
z} = -\frac{\partial p}{\partial z}
\end{eqnarray}
\begin{equation} \label{B.4}
\frac{1}{r}\frac{\partial}{\partial r}(r v_r) + \frac{\partial
v_z}{\partial z} =0
\end{equation}
We seek  a solution of the system  (\ref{cylindr}), (\ref{B.4}) in
the linear form
\begin{equation}
v_{\phi} = \omega(t) r\,, \quad v_r = a(t) r\,,\quad v_z = b(t) z
\label{B.5}
\end{equation}
The  corresponding pressure following from (\ref{B.5})  must be
$$
p(r,z,t) = \frac{P_1(t)}{2} r^2 + \frac{P_2(t)}{2} z^2
$$
The equation (\ref{B.4})  then implies a relation between $a$ and
$b$:
\begin{equation}
2a + b = 0\,.  \label{B.6}
\end{equation}
This means  the fluid volume conservation. Indeed, we consider a
cylindrical drop with a radius $R(t)$ and a length $Z(t)$ at an
instant $t$. Then it follows from (\ref{B.5}) that
$$
\dot{R} = a(t) R\,,\quad \dot{Z} = b(t)Z\,,
$$
Integrating these equations, we make sure that the cylinder volume
conserves:
$$
\pi R(t)^2 Z(t) = \pi R_0^2Z_0 \exp\int_0^t\left(2a(t_1) +
b(t_1)\right) dt_1 = \pi R_0^2 Z_0.
$$

Combining   (\ref{B.5}) with (\ref{cylindr}), we obtain a system
of ordinary differential equations:
\begin{eqnarray}
\dot{a} + a^2 - \omega^2 = - P_1 \nonumber \\
\dot{\omega} + 2 a \omega = 0    \label{B.7}\\
\dot{b} + b^2 = - P_2 \nonumber
\end{eqnarray}
Differentiating the second equation of system (\ref{B.7}) and
substituting other equations, we obtain
\begin{equation}\label{cilom}
\ddot{\omega} = -P_2(t)\, \omega
\end{equation}
The function  $P_2(t)$ has a meaning of pressure fall along the
cylinder axis. Let us suppose that it is determined by large-scale pressure
pulsations in turbulent flow. Hence, it is a complicated function of time; we assume
that its time average is zero.
 Then the time intervals when $P_2(t)>0$ and $P_2(t)<0$
are equally probable. However, at $P_2(t)>0$ the function
$\omega(t)$ oscillates, the oscillation amplitude changing weakly.
On the contrary, at $P_2(t)<0$, the function $\omega(t)$ grows
exponentially. It is clear that the value of $\omega$ grows on
average. We note also that from (\ref{B.7}) and (\ref{B.6})
follows the proportionality of  $\omega$ to the cylinder length
$Z$. Hence, such growth of $\omega$ means stretching  the
cylinder. We also note that, despite the nonlinearity of the
initial system (\ref{cylindr}), (\ref{B.4}), the final equation
(\ref{cilom}) is linear. The nonlinearity acts only in the
directions transversal to the cylinder axis.

This example illustrates the behavior of a drop with large
vorticity in a turbulent flow: it stretches out forming a
filament, and the vorticity  continues to increase. In \cite{ZSIG}
we have analyzed the general equations (\ref{hd}) in terms of
Lagrangian variables. We have shown that the development of
filaments occurs  
in a way analogous to that in the considered example. We have
found a linear equation analogous to (\ref{cilom}) describing the
growth of vorticity in the accompanying reference frame:
\begin{equation}\label{ronk}
\ddot{x}_n = - \rho_{nk}x_k\,,\qquad x_i=\omega_i\,,\qquad
\rho_{nk}=\nabla_n\nabla_k P
\end{equation}
Here $x_i$ and $P$ are the components of 
$\bomega$ and the pressure  in the 
Lagrangian reference frame. The matrix $\rho_{nk}$ is generated by
large-scale pulsations. In order to obtain the statistical
properties of the flow, it must be considered as a random
quantity. Hence, (\ref{ronk}) is a stochastic equations set.


 Instead of three equations of the second order,
we consider a set of six first-order equations:
\begin{equation}\label{xy}
    \dot{x_i}=y_i \, \qquad \dot{y_i}=-\rho_{ij} x_j
\end{equation}
We introduce the joint probability density
\begin{equation}   \label{sovmestxy}
f(t,{\bf x},{\bf y}) = \langle \delta({\bf x}-{\bf
x}(t))\delta({\bf y}-{\bf y}(t))\rangle  \ .
\end{equation}
Here ${\bf x}(t)$ and ${\bf y}(t)$ are the solutions of (\ref{xy})
at the given realization of $\rho_{ij}$ and initial conditions;
the average is taken over the ensemble of all possible
realizations. The variables $\bf x$ and $\bf y$ are independent.

Let the flow be  locally homogeneous and isotropic. Then, assuming
statistical independency of different components of the matrix
$\rho_{ij}$, we obtain from (\ref{xy}) the Fokker-Planck equation
for $f$ (see \cite{ZSIG}):
\begin{equation}\label{f}
 \frac{\d f}{\d t} + { y_k} \frac{ \d f}{\d { x_k}} = \left[x^2\frac{ \d^2 f}{ \d {\bf y}^2 }+
     \left( { x_k} \frac{\d}{\d y_k} \right)^2 f\right] \ ,
\end{equation}
where  $t$ is normalized by  the characteristic time of the
probability density change that is of the order of the correlation
time.

For what follows we need  an asymptotic time dependence of the
moments $<x^{2n}>$. To find it we consider the invariant moments
of the even order $2P$:
$$
M_P(n,m,k)=\left< x^{2n} y^{2m} ({\bf x y})^k\right> \equiv \int
x^{2n} y^{2m} ({\bf x y})^k f d{\bf x} d{\bf y} \,,\qquad\qquad
n+m+k=P
$$
Integrating (\ref{f}) with appropriate weights, we obtain for
these moments a closed set of linear differential equations
$$
\frac{d}{dt}M_P(n,m,k) = 2n M_P(n-1,m,k+1) + 2m(4k+2m+2)
M_P(n+1,m-1,k)
$$
$$
+4m(m-1) M_P(n,m-2,k+2) +k M_P(n,m+1,k-1) + 2k(k-1) M_P(n+2,m,k-2)
$$
The  number $Q$ of equations in the set   is equal to a number of
combinations $(n,m,k)$ such that $n+m+k=P$:
$$
Q=\frac12 (P+1)(P+2)
$$
Thus, the evolution of the $2P$-order invariant moments is
described by the set 
of $Q$ linear differential
equations. 
As $t\to\infty$, the solutions increase exponentially, in
particular $\langle x^{2n} \rangle \propto exp(\Lambda_{2n}t)$,
$\Lambda_{2n} $ being the maximal root of the corresponding
characteristic equation. Since $ <x^{2n}> $ is positive,
the solution must not oscillate, and the root must be real. To
find the exponents, we calculate the determinants  and solve the
characteristic (algebraic) equations numerically. The first
sixteen values $\Lambda_n$ are
\begin{equation}\label{L_n}
\begin{array}{llll}
\Lambda_2=2.52\,,\qquad & \Lambda_4=6.12\,,\qquad &
\Lambda_6=10.43\,,\qquad & \Lambda_8=15.25\,,\qquad \\
 \Lambda_{10}=20.48\,,\qquad & \Lambda_{12}=26\,,\qquad &
 \Lambda_{14}= 32.03\,,\qquad & \Lambda_{16}=38.25\,\qquad \\
\Lambda_{18}=44.73\,,\qquad & \Lambda_{20}=51.46\,,\qquad &
\Lambda_{22} = 58.42 \,,\qquad & \Lambda_{24}=65.58 \\
\Lambda_{26}=72.95\,,\qquad & \Lambda_{28}=80.52\,,\qquad & \Lambda_{30}=88.26
 & \Lambda_{32}=96.16
\end{array}
\end{equation}
To summarize, we cite  \cite{ZSIG} to list the main properties of
the equation (\ref{f}) concerning the moments:

1. All $n$-order moments of  $x_k$ and $y_j$  are connected by a
set of first-order linear differential equations. Hence, it is
possible to evaluate the moments of any order.

2. The even moments grow exponentially. Independently of the
initial conditions, the function $f$ at large values of $t$
depends only on the moduli $x$ and $y$ and on the cosine of the
angle between the vectors $\mu=({\bf x},{\bf y})/xy$.

3. The higher even moments grow faster than the lower ones.

The properties 2  and 3 express the presence of intermittency. For
example, for large values of $t$ we have  $<x^{2n}> \gg <x^2>^n$.

\section{Lagrangian structure functions}

As shown in Section 2, regions of large vorticity in a turbulent
flow take the form of filaments stretched along the vorticity
direction. 
\begin{figure}[h]
\hspace{4.5cm}
\includegraphics[width=7cm]{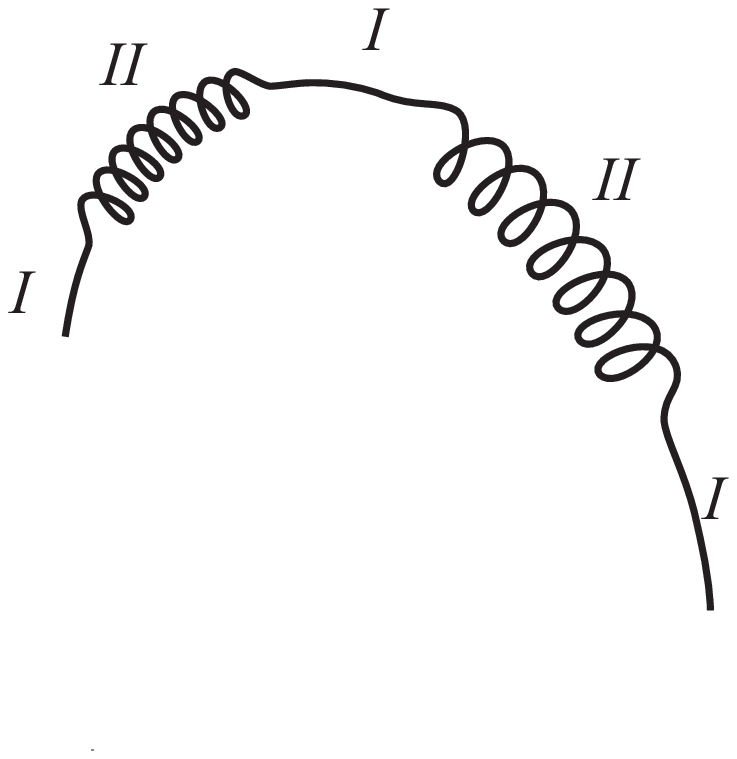}

{\hspace{5cm}\small Fig.1.   A typical trajectory of a test
particle. }
\end{figure}

Then a typical trajectory in the flow demonstrates two types of behavior 
shown in Fig. 1. In the region I the vorticity is not very large,
and the trajectory is smooth, the characteristic time of velocity
change is of the order of the correlation time $\tau_c$. In the
region II the vorticity is large, the test particle oscillates
with frequency $\omega$. These second regions give the main
contribution to correlation functions. Indeed, the second order
Lagrangian structure function $ K_2(\tau)$ is proportional to
$\tau^2$ in "smooth" parts of trajectory, and we shall see below
that in the "oscillating" parts it is proportional to $\tau$.

So, we consider the regions with large $\omega$. We choose a
cylindric frame with the  origin in the local maximum of $\omega$,
the axis $z$ directed along $\omega$. Since there is fast rotation
around the $z$ axis  near the origin, the flow does not depend on
the angle $\phi$. Then in the linear approximation we obtain the
expression (\ref{B.5}) which returns us to the example considered
in Section 1:
\begin{equation}\label{tr}
v_{\phi} = \omega(t) r\,, \quad v_r = a(t) r\,,\quad v_z = b(t) z
\end{equation}
Conservation laws for mass and for angular momentum along the $z$
axis take the form (see (\ref{B.6}), (\ref{B.7}))
$$
2a + b = 0 \ ,
$$
$$ 
\dot\omega = -2 a(t) \omega
$$
From   (\ref{tr}) we find
\begin{equation} \label{tr1}
\frac{d r}{dt}= a(t) r\,,\qquad \frac{d z}{dt}=b(t) z\,,\qquad
\frac{d\phi}{dt}=  \omega(t)
\end{equation}
The parameters  $\omega, a, b$ change slowly, with characteristic
time $\tau_c$. To the contrary, the oscillations are fast,
$\omega\tau_c \gg 1$. Hence, with accuracy $\tau/\tau_c$ we have
\begin{equation} \label{deltav}
\left| \delta {\bf v} \right|= \left| {\bf v} (t+\tau) - {\bf v}
(t) \right| = 2 v_{\phi} \sin \frac{\delta \phi}2 = 2 r \omega
\sin \frac{\omega \tau}2
\end{equation}
To calculate the
correlation functions of degree $n$ we must raise an absolute
value of $\delta {\bf v}$        
to the $n$-th power and take an ensemble average with the
probability distribution function.   


We first consider the pair correlation function
$$
K_2(\tau) = \langle (v(t+\tau)-v(t))^2 \rangle
$$
In accordance with the above, it has the form
\begin{equation}\label{K2}
K_2(\tau) \propto \int P(\omega,t)\,\omega^2\,\sin^2(\omega\tau /
2)\,d\omega
\end{equation}
In \cite{ZSIG} we have  analyzed the solution of (\ref{f}) and the
integrated PDF $P(x)=\int f(x,y,\mu) x^2 d{\bf y} d\mu$ (here
$\mu$ is the cosine of the angle between $\bf x$ and ${\bf y}$).
We have found that $P(\omega)$ has an  intermediate  stationary
asymptote 
\begin{equation}\label{intF}
P(\omega)=  C  \omega^{-4}
\end{equation}
For small and for infinitely large values $\omega$, the PDF
remains non-stationary.

We see that the integral (\ref{K2}) converges for the stationary
PDF (\ref{intF}). Actually,
$$
K_2(\tau)\propto
\int_0^{\infty}\frac{\omega^2\,\sin^2(\omega\tau/2)}{\omega^4}\,d\omega
\propto \tau \int_0^{\infty}\frac{\sin^2(q)}{q^2}\,dq
$$
(The input of $\delta v_z^2$ is small -- proportional to $z^2$).

So, for  $\tau \ll \tau_{c}$ the second order structure function
along the Lagrangian trajectory is
\begin{equation}  \label{K2tau}
K_2(\tau)\propto \tau
\end{equation}
This result adjusts with the Kolmogorov's theory K41
\cite{Landau},\cite{Frisch}.

For higher-order structure functions $K_n(\tau)$ we have
\begin{equation}\label{n-th}
K_n(\tau) \propto \int_0^{\infty} P(\omega,t)
\omega^n|\sin(\omega\tau /2)|^n d\omega
\end{equation}
The integral with the stationary PDF diverges. Hence, the higher
structure functions are determined by the non-stationary part of
the PDF. To calculate $K_n(\tau)$, we use the moments
$$
\int_0^{\infty}P(\omega,t) \omega^{2n} d\omega \propto
e^{\Lambda_{2n}t}
$$
found in Section 2.  We expand the sine in (\ref{n-th}) into the
series and integrate each term. For time $t$ satisfying
$$
1 << t <<
\frac{\ln(\tau^{-2})}{\Lambda_{2n+2}-\Lambda_{2n}}\,,\qquad\qquad
\tau\to 0
$$
the second item is much less than the first one. We restrict
ourselves by the first term as $\tau \to 0$. Then
\begin{equation}\label{ser}
K_n(\tau) \propto \tau^n e^{\Lambda_{2n}t}
\end{equation}
Excluding $t$, we obtain
\begin{equation}  \label{chain}
\left(\frac{K_m(\tau)}{\tau^m}\right)^{1/\Lambda_{2m}}  =
\left(\frac{C_{n|m} K_n(\tau)}{\tau^n}\right)^{1/\Lambda_{2n}}
\end{equation}
Here $C_{n|m}$ are constants not depending on $\tau$. This
relation connects the structure functions of different orders.
Choosing $m=2$ and taking (\ref{K2tau}) into account, we find
$$K_n\propto \tau^{\zeta_n}\ ,$$
where
\begin{equation}\label{sigma}
\zeta_n= n-\frac{\Lambda_{2n}}{\Lambda_4}\,
\end{equation}
Using (\ref{L_n}) we get first twelve scaling exponents $\zeta_n$:
$$
\begin{array}{llll}
\zeta_1 = 0.59\,, \qquad & \zeta_2 = 1\,,\qquad & \zeta_3 = 1.3\,,
\qquad & \zeta_4 = 1.51\,, \\  \zeta_5 = 1.65\,,\qquad & \zeta_6 =
1.75 \,, \qquad & \zeta_7 = 1.77\,,\qquad & \zeta_8 = 1.75\,, \\
\zeta_9 = 1.69\,, \qquad &
 \zeta_{10} = 1.59\,,\qquad & \zeta_{11}=1.45\,,\qquad & \zeta_{12}=1.28
 \end{array}
$$

Actually, the stationary structure function (\ref{K2tau}) is not necessary to
determine  the scaling exponents. The reason is that the
parameters $\Lambda_{2n}$ grow fast as a function of $n$.  For
example, we take $m=16$. Using (\ref{chain}) with
$\Lambda_{32}=96.16$,  we then obtain
$$
K_n(\tau) \propto
\tau^{n-16\Lambda_{2n}/\Lambda_{32}}\,K_{16}^{\Lambda_{2n}/\Lambda_{32}}
$$
For small $n$ the ratio $\Lambda_{2n}/\Lambda_{32}$ is small
enough, and we can put $K_{16}^{\Lambda_{2n}/\Lambda_{32}} \approx
1$. For $n=2$ we then find
$K_2(\tau)\propto\tau^{0.98}\approx\tau$. The result is very close
to that obtained with the stationary PDF.
We note that this relation is derived now from (\ref{chain}) only.
This demonstrates the consistency between the stationary and
non-stationary solutions of the equation (\ref{f}). Taking other
values of $n$, we get in the same way
$$
\zeta'_1 = 0.58\,,\qquad\qquad \zeta'_2 = 0.98\,,\qquad\qquad
\zeta'_3 = 1.26
$$
$$
\zeta'_4 = 1.46\,,\qquad\qquad \zeta'_5 = 1.59\,,\qquad\qquad
\zeta'_6 = 1.67
$$

We recall that in the derivation of (\ref{chain}) we neglected the
rest of the sine expansion in (\ref{n-th}) as $\tau \to 0$. To
justify this approximation, we calculate (\ref{n-th}) with
complete sine expansion, using the moments of $\omega$. We then
express $e^t$ through  $K_{16}$ using (\ref{ser}).
For example, for $K_2$ we have
\begin{equation}\label{rjad}
K_2 =\frac{1}{4} \tau^2 e^{\Lambda_4t} -\frac{1}{48} \tau^4 e^{\Lambda_6t} +
\frac{1}{1440}\tau^6 e^{\Lambda_8t} + \, ... 
\end{equation}
$$
= \frac{1}{4}\tau^{0.98}K_{16}^{0.06} -\frac{1}{48} \tau^{2.26} K_{16}^{0.11} +
\frac{1}{1440}\tau^{3.46} K_{16}^{0.16} + \, ...
$$
We see that as $\tau\to 0$ the series converges rapidly, and the
rest is small with respect to the first term. The same is correct
for the structure functions of higher orders.

We note also  that the relation
(\ref{chain}) can be rewritten as
\begin{equation}\label{const}
 C_{n|m} =
 \frac{K_m^{\Lambda_{2n}/\Lambda_{2m}}}{K_{n}}\,\tau^{n-m\frac{\Lambda_{2n}}{\Lambda_{2m}}}
\end{equation}
According to our theory,  these combinations of structure
functions must  be constant, i.e. they must  not depend on $\tau$.

\section{Comparison of the theory with the experiment.}

In this Section we compare the predictions of our theory with  the
results of recent experiments of the Bodenschatz group
\cite{Bod1},\cite{Bod2}. In these experiments turbulence was
generated by two counter-rotating disks. Different regimes of the
facility allowed to make the measurements for different local
Reynolds numbers $R_{\lambda}$. The Lagrangian test particles used
to trace the flow  had the size comparable or smaller than the
Kolmogorov scale $\eta$ for all tested Reynolds numbers. The
motion of particles was tracked in a subvolume of about $10^{-4}$
of the whole turbulent volume, in the center of the tank where the
effects of the mean velocity were negligible. Thus, this region
seems to us optimal for investigation of locally homogenous
turbulence.

The measurements performed in \cite{Bod1} showed directly the
presence of stretched structures -- vortex filaments. It was shown
that near to these structures the absolute values of particle
accelerations were maximal. The probability distribution of these
accelerations was found to be significantly non-gaussian in the
region of their large values. Thus, general picture of  vortex
structures in turbulence corresponds to that in our theory
\cite{ZSIG}.

\begin{center}
Table 1.  {\small Values of the scaling exponents normalized by
$\zeta_2$. The experimental results for different local Reynolds
numbers are cited from \cite{Bod2} }\\
\begin{tabular}{l l l l l}
  & & \\ \hline
$R$      & 200    & 690    & 815   & Theory \\
\hline
$\zeta_1/\zeta_2$ &  $0.59\pm 0.02$ & $0.58\pm 0.05$ & $0.58\pm 0.12$ &  0.59 \\
$\zeta_3/\zeta_2$ &  $1.24\pm 0.03$ & $1.28\pm 0.14$ & $1.28\pm 0.30$ &  1.3  \\
$\zeta_4/\zeta_2$ &  $1.35\pm 0.04$ & $1.47\pm 0.18$ & $1.47\pm 0.38$ &  1.51 \\
$\zeta_5/\zeta_2$ &  $1.39\pm 0.07$ & $1.61\pm 0.21$ & $1.59\pm 0.46$ &  1.65 \\
$\zeta_6/\zeta_2$ &  $1.40\pm 0.08$ & $1.73\pm 0.25$ & $1.66\pm 0.53$ &  1.75 \\
$\zeta_7/\zeta_2$ &  $1.39\pm 0.09$ & $1.83\pm 0.28$ & $1.67\pm 0.60$ &  1.77 \\
$\zeta_8/\zeta_2$ &  $1.40\pm 0.10$ & $1.92\pm 0.32$ & $1.65\pm 0.66$ &  1.75 \\
$\zeta_9/\zeta_2$ &  $1.42\pm 0.11$ & $1.97\pm 0.35$ & $1.61\pm 0.73$ &  1.69 \\
$\zeta_{10}/\zeta_2$ &  $1.46\pm 0.12$ & $1.98\pm0.38$ & $1.57\pm 0.80$ & 1.59 \\
\hline
\end{tabular}
\end{center}
\vspace{0.5cm}

We make the quantitative comparison of the  Lagrangian structure
functions obtained in our theory  with the results of the
experiment \cite{Bod2}. Table 1 and Fig. 2 show the Lagrangian
scaling exponents $\zeta_n$ normalized by $\zeta_2$, up to the
10th order. The first three columns of Table 1 represent the
values measured in \cite{Bod2} with different local Reynolds
numbers $R_{\lambda}$. The last column contains our theory
prediction (\ref{sigma}). The theory is in excellent agreement
with the experimental data. Moreover, the agreement is better for
larger values of $R_{\lambda}$. This corresponds to the fact that
the theory is constructed in the limit $R\to\infty$. Also, both in
the theory and in the experiment with the most value of
$R_{\lambda}$, the scaling exponents grow up to the 7-th order,
and then decrease.
 We note that the theory predicts also the
scaling exponents of higher orders that are not measured yet.

\begin{figure}[h]
\hspace{1.5cm}
\includegraphics[width=12cm]{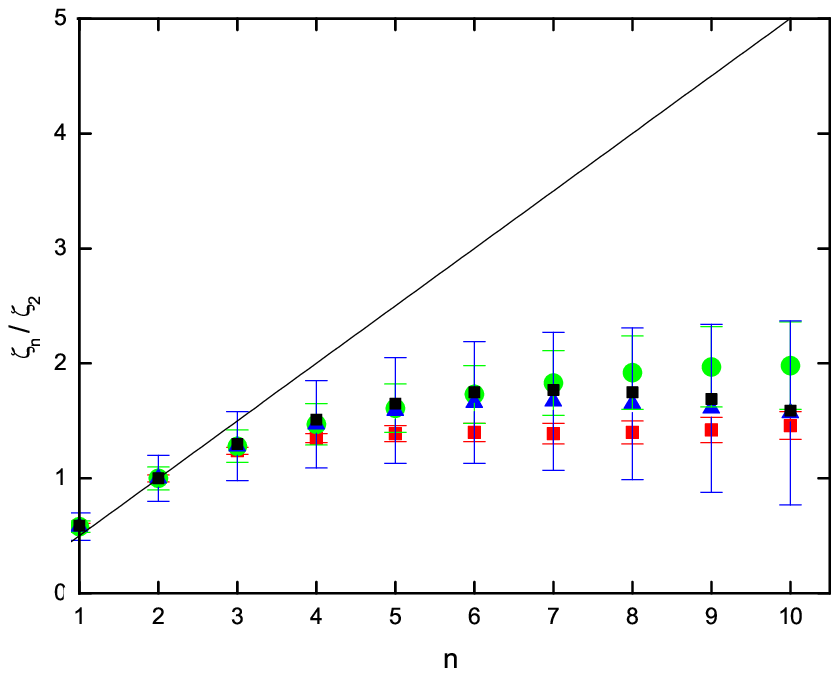}

{\small Fig.2 Scaling exponents $\zeta_n$ normalized by $\zeta_2$
as a function of order. Small black squares denote the prediction
of our theory, other symbols show the experimental results
\cite{Bod2} for different local Reynolds numbers:
$R_{\lambda}=200$ (red squares), $R_{\lambda}=690$ (green
circles), and $R_{\lambda}=815$ (blue triangles). The solid line
corresponds to the Kolmogorov theory K41. Strong departure from
the K41 prediction of both the experiment \cite{Bod2} and our
theory is the manifestation  of intermittency. }
\end{figure}

Table 1 and Fig. 2 also demonstrate the growth of dispersion of
the measured scaling exponents as a function of  order and of $R_{\lambda}$.
This is in accord with the presented theory, which is based on the
notion that  filaments (narrow regions with very high vorticity)
make the most contribution to structure functions. The dispersion
behavior is then the result of intermittency. The higher is the
order, the more important become very high peaks, the more seldom
they occur. Hence, for a given sample the dispersion increases
with the moment order.
On the other hand, the real height of  peaks is limited by
viscosity.\footnote{Presumably, the viscosity constrains the value
of $m$ in (\ref{chain}). We choose $m=16$ as an upper limit
because the ratio of the large-scale characteristic time $\tau_c$
to the viscous Kolmogorov time $\tau_{\eta}$ is about 100. Since
the time of change of $K_{16}$ is $\Lambda_{32}^{-1}\approx
0.01\tau_c$, it lies near the boundary between the inertial  and
the viscous ranges.} As viscosity decreases,  the size of
a sample needed to determine the statistical moments 
increases. This property of statistical systems is demonstrated,
in particular, in \cite{Sokol1}. 

In addition to analysis of the scaling exponents, we also compare
(\ref{chain}) 
with direct experimental data \cite{Bod2} for dependence of the
high-order Lagrangian structure functions of $\tau$. For that we
choose the form (\ref{const}) of the equation (\ref{chain}) to
check if the combinations $C_{n|m}$ of any two structure functions
are constant. The results for $C_{10|9}$ and $C_{10|8}$ are
presented in Fig.3. We see that even for the highest of the
measured orders, the values $C$ are constant up to the accuracy of
the available experimental graph.

\begin{figure}[h]
\hspace{1.5cm}
\includegraphics[width=12cm]{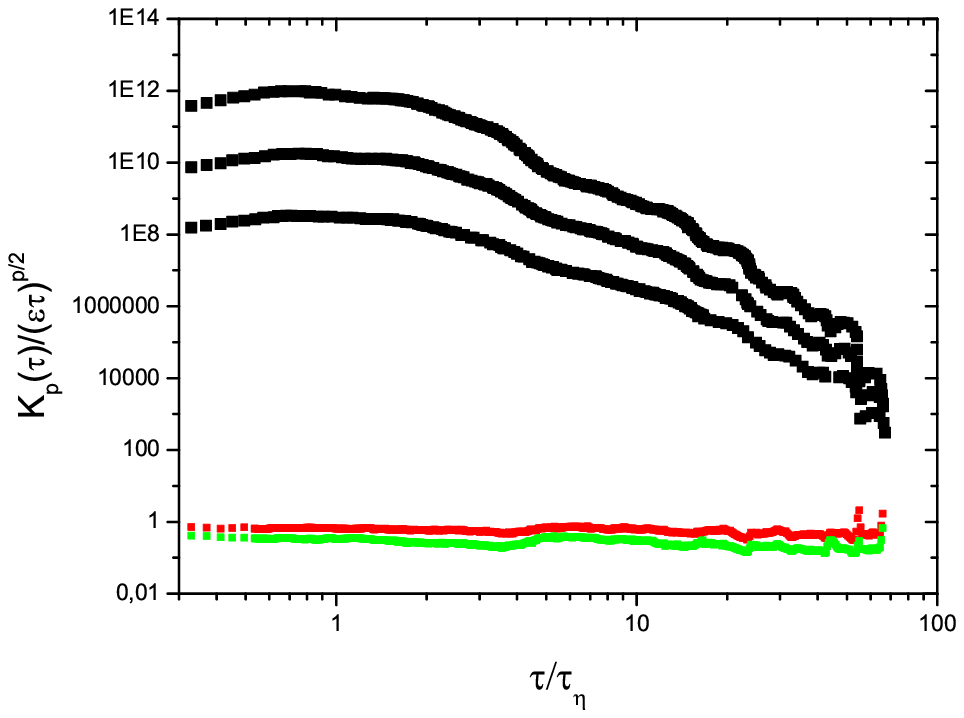}

{\small Fig.3.  Lagrangian structure functions of the orders 8,9,
10 (black curves from the bottom to the top)  normalized by the
K41 predictions ($R_{\lambda}=690$): cited from \cite{Bod2}. Red
and green curves denote the functions $C_{10|9}$ and $C_{10|8}$,
respectively.}
\end{figure}

We stress that  no adjustable parameters were used  in calculating
both the scaling exponents (Table 1, Fig.2) and  the combination
of the Lagrangian structure functions (Fig.3). Thus, the
coincidence between the theoretical predictions and the
experimental results is wonderful. One can suppose that a
significant role belongs to a happy  choice  of parameters of the
experiment: very homogeneous flow in the subvolume where the
measurements were performed, and the size of test particles
coinciding with the Kolmogorov scale.
 Due to this the high frequency noise is
damped in the measurements, and the experimental data  describe
the inertial range only. Exactly this range is studied by the
theory.

\section{Universal scaling for Lagrangian structure functions of
position.}

The Lagrangian approach gives a possibility to study not only  the
statistical properties of velocities, but also that of positions
of the particles. This allows to introduce a new notion in  the
turbulence theory. Define  the Lagrangian structure functions of
positions of two particles as
$$
R_n(\tau)=\left<|{\bf r}(t+\tau)-{\bf r}(t)|^n\right>
$$
As in the case of velocity structure functions, the main
contribution in $R_n$ comes from the regions with large vorticity.
Integrating (\ref{tr1}) and calculating $|\delta {\bf r}|$ by
analogy with (\ref{deltav}), we obtain
$$
R_n(\tau) \propto \int P(\omega)
\sin^n(\omega\tau/2)\,d\omega\propto
\int\frac{\sin^n(\omega\tau/2)}{\omega^4}\,d\omega
$$
If  $n>3$ the integral  with the stationary PDF (\ref{intF})
converges. Hence, the high-order structure functions of positions
are determined by the stationary part of the PDF. We note that in
the asymptotics all the scaling exponents are equal:
$$ R_n(\tau)\propto
\int\frac{\sin^n(\omega\tau/2)}{\omega^4}\,d\omega\, \propto
\tau^3 \,,\qquad\,n>3
$$
In other words, there is a universal  scaling for all $R_n$,
$n>3$.
 This universal scaling, if it would be observed in experiments,
 can be  an evidence  for the determining influence of the filaments  on the
behavior of the structure functions in the inertial range.

\section{Conclusion}

To conclude, we list the main results of the paper.
\begin{enumerate}
\item  The Lagrangian velocity structure functions of high orders
are derived in the inertial interval of fully developed liquid
turbulence. 
\item The comparison of the theory predictions with experimental results
shows an excellent agreement.
\item New experimentally measurable values -- the Lagrangian position structure functions
are introduced and calculated.  Their scaling exponents are shown
to be independent on the order of the function for any $n>3$.
\end{enumerate}
We emphasize that no adjustable parameters were used for the
scaling exponents.
Also  the power-law dependence 
of the Lagrangian structure functions was not suggested in the
model. All the obtained relations are the consequences of the
equation (\ref{f}) derived directly from the Navier-Stokes
equation in the inertial interval.

\end{document}